\documentclass[%
 reprint,
superscriptaddress,
showpacs,
 amsmath,amssymb,
 aps,
 pra,
 floats,
]{revtex4-1}

\usepackage[latin1]{inputenc}
\usepackage{graphicx}
\usepackage{dcolumn}
\usepackage{bm}
\usepackage{units}
\usepackage{color}

\newcommand{\ket}[1]{ \left|#1\right\rangle}
\begin{document}

\preprint{APS/123-QED}

\title{Optical wire trap for cold neutral atoms}

\author{Philipp Schneeweiss, Fam Le Kien, and Arno Rauschenbeutel}
\affiliation{%
 Vienna Center for Quantum Science and Technology,\\
 TU Wien -- Atominstitut, Stadionallee 2, 1020 Vienna, Austria
}%

\date{\today}

\begin{abstract}
We propose a trap for cold neutral atoms using a fictitious magnetic field induced by a nanofiber-guided light field. In close analogy to magnetic side-guide wire traps realized with current-carrying wires, a trapping potential can be formed when applying a homogeneous magnetic bias field perpendicular to the fiber axis. We discuss this scheme in detail for laser-cooled cesium atoms and find trap depths and trap frequencies comparable to the two-color nanofiber-based trapping scheme but with one order of magnitude lower powers of the trapping laser field. Moreover, the proposed scheme allows one to bring the atoms closer to the nanofiber surface, thereby enabling efficient optical interfacing of the atoms with additional light fields. Specifically, optical depths per atom, $\sigma_0/A_{\rm eff}$, of more than 0.4 are predicted, making this system eligible for nanofiber-based nonlinear and quantum optics experiments.

\end{abstract}
\maketitle



Trapping and optically interfacing cold neutral atoms in the near field of nanophotonic structures have attracted considerable attention in recent years~\cite{Vetsch10,Goban12,Thompson13b}. A successful and highly promising approach in this endeavor relies on the use of optical dipole forces of a blue- and a red-detuned nanofiber-guided light field in order to form a so-called two-color trap~\cite{Dowling96,LeKien04}. This scheme has been demonstrated experimentally for laser-cooled cesium, storing the atoms at about 200~nm above the nanofiber surface~\cite{Vetsch10,Goban12}. Other types of nanofiber-based traps for cold atoms have been discussed theoretically, relying on, e.g.,~the combination of an attractive potential of a red-detuned field and the repulsive potential of the centrifugal force~\cite{Balykin04}, the interference of higher-order modes~\cite{Sague08,ArXiv_Phelan13}, a diffracted laser field impinging perpendicularly to the fiber~\cite{LeKien09c}, or by modifying the two-color scheme to form a helical trapping potential~\cite{Reitz12}.

In this paper, we propose a nanofiber-based trap for cold neutral atoms that relies on the Zeeman-state-dependent energy level shift induced by a nanofiber-guided light field. Our work has close conceptional ties with optically induced hybrid traps on atom chips~\cite{Yang08} and is focused on exploiting the advantages offered by optical nanofibers. Specifically, our trapping scheme features a lower required power of the guided trapping light compared to the two-color trap as well as a smaller distance between the atoms and the fiber surface. This enables a better coupling of the trapped atoms to additional light fields. Moreover, the proposed scheme is closely related to magnetic wire traps, allowing one to benefit from the well-established methods developed in this field. 

\begin{figure}
	\includegraphics{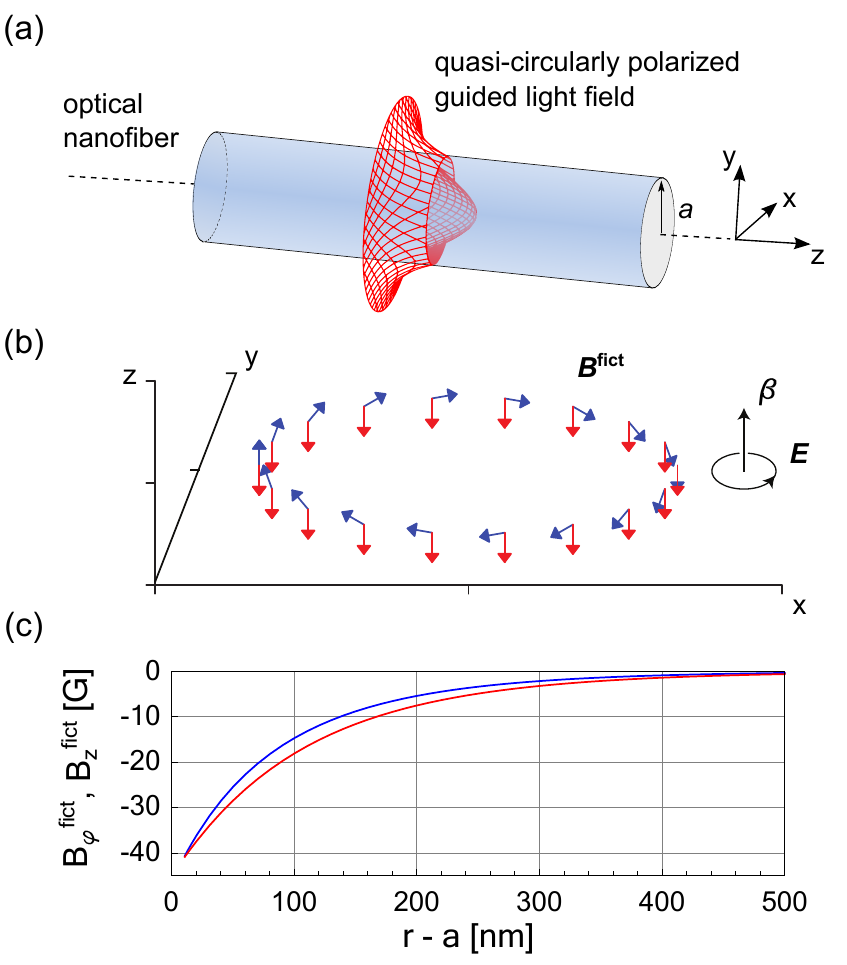}
	\caption{(a) Schematic intensity distribution of a nanofiber-guided light field propagating in the $+z$ direction with quasi-circular polarization. (b) Schematic vector profile of the fictitious magnetic field $\mathbf{B}^{\mathrm{fict}}$ induced by the light field shown in (a) with counter-clockwise polarization. A cesium atom in the $6S_{1/2}$ ground state is assumed and the wavelength of the guided field lies between the cesium D-lines. (c) Radial dependence of the components $B^{\mathrm{fict}}_{\varphi}$ (blue curve) and $B^{\mathrm{fict}}_{z}$ (red curve) of the fictitious magnetic field. Calculations are for a guided field at $\lambda_{\rm trap}=880.25$~nm of optical power $P=1.2$~mW and a radius of the nanofiber of $a=230$~nm.}
	\label{fig:principle} 
\end{figure}

We consider a cylindrical dielectric waveguide of refractive index $n$ and radius $a$ surrounded by vacuum. For a sufficiently small radius $a$, typically a few hundred nanometers, this system acts as a single-mode wave\-guide for light. The only sustained mode of this optical nanofiber is the hybrid HE$_{11}$ mode~\cite{Snyder83}. Light propagating in optical nanofibers is strongly guided and, hence, exhibits a significant longitudinal polarization component. In addition, a large fraction of the total optical power $P$ of the guided field propagates outside of the nanofiber in the form of an evanescent wave~\cite{LeKien04b}. We sketch a typical intensity distribution for a quasi-circularly polarized nanofiber-guided mode in Fig.~\ref{fig:principle}(a).

Generally, the energy levels of an atom exposed to a light field are modified due to the ac-Stark effect. For alkali atoms in their electronic ground state, this light shift has two contributions arising from the scalar and vector polarizabilities of the atom. The scalar light shift is the same for all Zeeman states within one hyperfine manifold. Here, we make this contribution vanish by using a nanofiber-guided trapping field at a ``tune-out'' wavelength~\cite{Arora11}. For $^{133}$Cs atoms, this is achieved at $\lambda_{\rm trap}=880.25$~nm, which is in between the D1 line ($\approx 894$~nm) and the D2 line ($\approx 852$~nm) such that their contributions to the scalar shift exactly cancel. The remaining effect of the nanofiber-guided light field on the atomic levels arises from the vector light shift only. The latter depends on the magnetic quantum number $m_F$ and can be expressed as the effect of a light-induced fictitious magnetic field~\cite{Cohen-Tannoudji72,LeKien13a}
\begin{equation}\label{eq:Bfict} 
\mathbf{B}^{\mathrm{fict}}
=\frac{\alpha^v_{nJF}}{8\mu_Bg_{nJF}F} i[\boldsymbol{\mathcal{E}}^*\times\boldsymbol{\mathcal{E}}]~.
\end{equation}
Here, $\mu_B$ is the Bohr magneton, $g_{nJF}$ is the Land\'e factor for the hyperfine level $|nJF\rangle$, $\alpha^v_{nJF}$ is the conventional vector polarizability of the particular hyperfine level, and $\boldsymbol{\mathcal{E}}$ is the positive-frequency electric field envelope for a light field, which is related to the electric field $\mathbf{E}$ by $\mathbf{E}=1/2~(\boldsymbol{\mathcal{E}}e^{-i\omega t}+\mathrm{c.c.})$. The magnitude of the light-induced fictitious magnetic field $\mathbf{B}^{\mathrm{fict}}$ is proportional to the intensity of the light field, and the direction of $\mathbf{B}^{\mathrm{fict}}$ is determined by the sign of $\alpha^v_{nJF}$ and by the cross product $i[\boldsymbol{\mathcal{E}}^*\times\boldsymbol{\mathcal{E}}]$. If the light field exhibits a polarization with nonzero ellipticity, the fictitious magnetic field points in a direction perpendicular to the plane defined by the circulating electric field vector. For linearly polarized light, $B^{\mathrm{fict}}=0$.

In Fig.~\ref{fig:principle}(b), we show schematically the fictitious magnetic field $\mathbf{B}^{\mathrm{fict}}$ induced by a quasi-circularly counter-clockwise polarized light field guided in the HE$_{11}$ mode~\cite{ArXiv_LeKien13c}. The fictitious field $\mathbf{B}^{\mathrm{fict}}$ can be decomposed into two components, azimuthal and axial. The azimuthal ($\varphi$-) component (blue arrows) has the same orientation as the magnetic field around a current-carrying wire. The axial ($z$-) component of $\mathbf{B}^{\mathrm{fict}}$ (red arrows) is parallel to the direction of propagation of the guided field. This component has no equivalent in current-carrying wires. Furthermore, the radial ($r$-) dependence of $\mathbf{B}^{\mathrm{fict}}$ differs from the $1/r$-behavior encountered in the case of current-carrying wires. In Fig.~\ref{fig:principle}(c), we plot the azimuthal and axial components of $\mathbf{B}^{\mathrm{fict}}$ as a function of the radial position. Analytically, $\mathbf{B}^{\mathrm{fict}}$ is given by
\begin{align}
\mathbf{B}^{\mathrm{fict}} &= \frac{\alpha^v_{nJF}}{4\mu_Bg_{{nJF}}F}[ 
\mathrm{Im}(\mathcal{E}_z\mathcal{E}_r^*)\hat{\boldsymbol{\varphi}}+\mathrm{Im}(\mathcal{E}_r\mathcal{E}_\varphi^*)\hat{\mathbf{z}}]~, \\
\boldsymbol{\mathcal{E}} &= (\mathcal{E}_r, \mathcal{E}_\varphi, \mathcal{E}_z) \nonumber\\
&=A (\boldsymbol{\hat r} e_r + \boldsymbol{\hat \varphi} e_\varphi + \boldsymbol{\hat z} e_z) \exp (i \beta z + i \varphi)~.
\label{eq:normalization}
\end{align}
Here, $(\boldsymbol{\hat r}, \boldsymbol{\hat \varphi}, \boldsymbol{\hat z})$ are the unit vectors and $\mathbf{e}(\mathbf{r})$ is the unnormalized mode-profile vector function of the electric part of the fundamental guided mode HE$_{11}$. The components of the latter are given by~\cite{Snyder83}
\begin{eqnarray}
e_{r}&=&i[(1-s)K_0(qr)+(1+s)K_2(qr) ],
\nonumber\\
e_{\varphi}&=&-[(1-s)K_0(qr)-(1+s)K_2(qr) ],
\nonumber\\
e_{z}&=& \frac{2q}{\beta}K_1(qr)~.
\end{eqnarray} 
The above equations for $\mathbf{e}(r)$ are valid for $r>a$, i.e.,~outside of the nanofiber. The notations $J_n$ and $K_n$ stand for the Bessel functions of the first kind and the modified Bessel functions of the second kind, respectively. The parameter $s$ is defined as $s=({1}/{h^2a^2}+{1}/{q^2a^2})/[{J_1^\prime (ha)}/{haJ_1(ha)}+{K_1^\prime (qa)}/{qaK_1(qa)}]$ with $h=(n_1^2k^2-\beta^2)^{1/2}$ and $q=(\beta^2-n_2^2k^2)^{1/2}$. Other parameters are the propagation constant $\beta$ of the guided field, and the free-space wave number $k$ of the field. The normalization constant $A$ in Eq.~(\ref{eq:normalization}) can be determined from the optical power of the nanofiber-guided light field.

For an atom in the electronic ground state, the fictitious magnetic field behaves in almost every respect like a real magnetic field~\cite{Cohen-Tannoudji72,LeKien13a,ArXiv_LeKien13c}. In particular, both fictitious and real magnetic fields are pseudo-vectors. Thus, the fictitious magnetic field $\mathbf{B}^{\mathrm{fict}}$ can be vector-added to any static real magnetic field $\mathbf{B}$, such that the atom is in total exposed to the effective magnetic field $\mathbf{B}^{\rm eff}=\mathbf{B}^{\mathrm{fict}}+\mathbf{B}$. This has been found in early work by Cohen-Tannoudji and Dupont-Roc~\cite{Cohen-Tannoudji72}, and further supported in various experiments, demonstrating, e.g., optically induced spin precession in an atomic beam~\cite{Zielonkowski98}.

Based on these considerations and in analogy to the side-guide traps using current-carrying wires~\cite{Frisch33,Fortagh98,Denschlag99b}, we propose to create an {\em optical wire trap} for neutral atoms. For this purpose, we apply a magnetic bias field $\mathbf{B}_{\rm bias}$ perpendicular to the axis of the nanofiber in Fig.~\ref{fig:principle}. Similar to conventional side-guide wire traps, low-field-seeking atoms can be confined in the $(x,y)$-plane around a line of minimal magnetic field that forms parallel to the nanofiber axis. The shift of the internal-state energy of a paramagnetic atom due to the effective magnetic field $\mathbf{B}^{\rm eff}$ is given by
\begin{eqnarray} \label{eq:generalMagneticInt}
U_{\rm mag} = - \boldsymbol{\mu} \cdot \mathbf{B}^{\rm eff}~,
\end{eqnarray}
where $\boldsymbol{\mu}$ is the magnetic moment of the atom. If $\boldsymbol{\mu}$ can adiabatically follow the direction of the local effective magnetic field while the atom moves within the potential, then Eq.~(\ref{eq:generalMagneticInt}) simplifies to~\cite{Fortagh07}
\begin{eqnarray} \label{eq:simpleMagneticInt}
U_{\rm mag} = \mu_B g_{nJF} m_F |\mathbf{B}^{\rm eff}|~.
\end{eqnarray}
For the trap presented here, the separation between the atom and the nanofiber is on the order of a few hundred nanometers and surface effects have to be taken into account. We include the latter into our calculations using a van~der~Waals potential of the form $U_{\rm vdW}=-C_3/(r-a)^3$,  where $C_3=5.6\times10^{49}$~Jm$^3$~\cite{LeKien04} for cesium. The total potential energy shift of the atom is then given by $U=U_{\rm mag}+U_{\rm vdW}$.

\begin{figure}
	\includegraphics{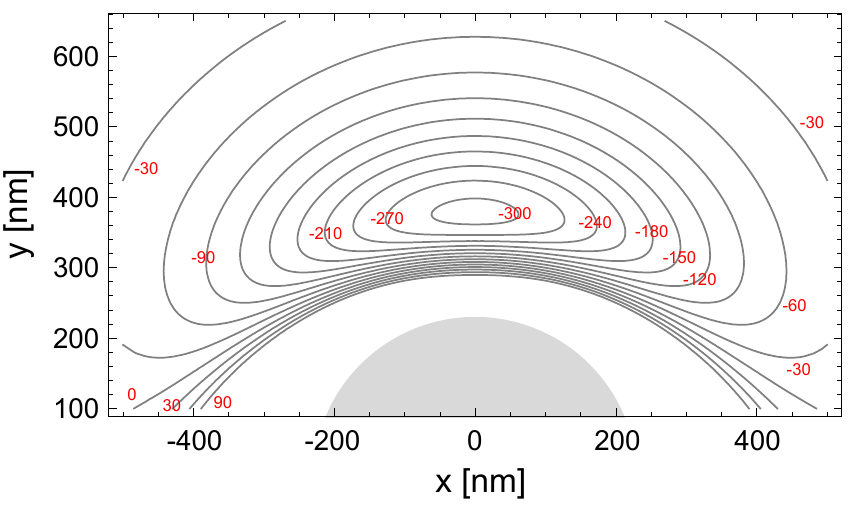}
	\caption{Contour plot of the potential energy of a cesium atom in the optical wire trap. The cross section of the nanofiber is marked as a solid gray disk at the bottom of the figure. The unit of the contour labels is $k_{\rm B}\cdot\mu$K where $k_{\rm B}$ is Boltzmann's constant. Parameters are as in Fig.~\ref{fig:principle}(c); a magnetic bias field $B_{\rm bias}=22$~G is applied in the $-x$-direction. The van~der~Waals potential induced by the nanofiber is taken into account as $U_{\rm vdW}=-C_3/(r-a)^3$ with $C_3=5.6\times10^{49}$~Jm$^3$.}
\label{fig:contour}
\end{figure}
We now discuss the characteristics of the optical wire trap. Specifically, we calculate the potential energy $U$ of a $^{133}$Cs atom in the state $\ket{6S_{1/2}, F=4,m_{\rm F}=4}$. Figure~\ref{fig:contour} shows the trapping potential calculated for a silica nanofiber with refractive index $n=1.45$, radius $a=230$~nm, a guided trap light field with a power of $1.2$~mW, and a homogeneous magnetic bias field of $22$~G, applied along the $-x$-direction. A local minimum of $U$ is formed about $150$~nm above the nanofiber surface at $(x=0,\;y_0\approx 150)$~nm. The depth of the potential $U_0/k_{\rm B} \approx 300$~$\mu$K in Fig.~\ref{fig:contour} is sufficient for storing laser-cooled cesium atoms. Note that, in contrast to conventional wire traps, the azimuthal component of the fictitious magnetic field and the applied bias field do \emph{not} completely cancel each other at the minimum of the trapping potential. This is connected to the presence of the spatially varying $z$-component of the fictitious magnetic field, $B^{\mathrm{fict}}_{z}$. When neglecting surface effects, the local minimum of the potential in the fiber transverse plane, $U(r,\varphi)$, can be determined from the condition $\nabla |\mathbf{B}^{\rm eff}(r,\varphi)|=0$. For a bias field oriented along the $-x$-direction, the minimum of the potential in the radial direction can be found be solving
\begin{align} 
0 = &~(B^{\rm fict}_\varphi(r) + B_{\rm bias} \sin \varphi) \partial_r B^{\rm fict}_\varphi(r) \nonumber \\
&+ B^{\rm fict}_z(r) \partial_r B^{\rm fict}_z(r)~.
\label{eq:radial}
\end{align}
As opposed to conventional wire traps, the term $\partial_r B^{\rm fict}_z(r)$ generally differs from zero in the case of the optical wire trap. 

\begin{figure}
\includegraphics{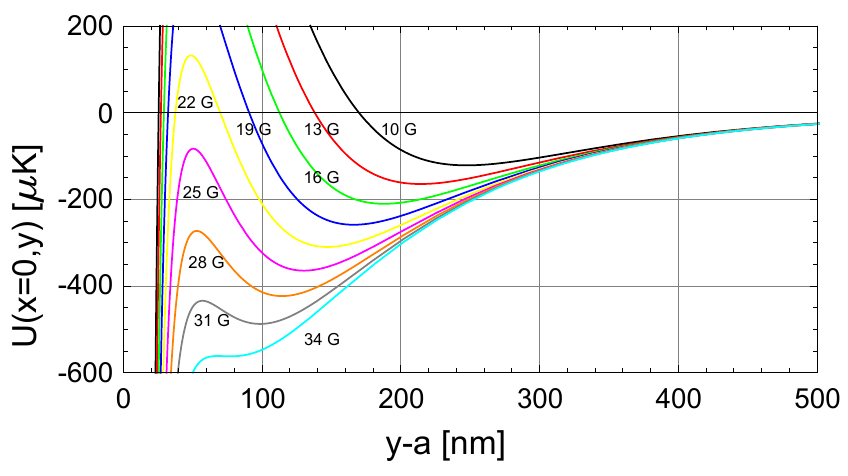}
	\caption{Trapping potential $U(x=0,y)$ for $P=1.2$~mW and $B_{\rm bias}=10 \ldots 34$~G. The black solid line corresponds to $B_{\rm bias}=10$~G, the cyan solid line to $B_{\rm bias}=34$~G, the increment is $3$~G. Other parameters are as in Fig.~\ref{fig:contour}.}
	\label{fig:TrapPos} 
\end{figure}
\begin{figure}
	\includegraphics{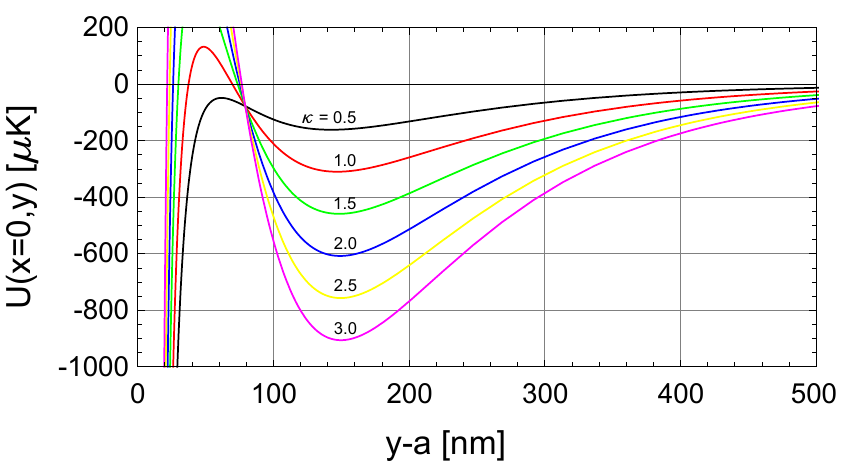}
	\caption{Trapping potential $U(x=0,y)$ for $P=\kappa\cdot1.2~{\rm mW}$ and $B_{\rm bias}=\kappa\cdot22~{\rm G}$ with $\kappa=\{0.5,~1,~1.5,~2,~2.5,~3\}$. The black solid line corresponds to $\kappa=0.5$, the magenta solid lines corresponds to $\kappa=3$. The trap depth is changing while the position of the trap minimum is approximately constant. Other parameters are as in Fig.~\ref{fig:contour}. The common crossing point of all potential lines results from the chosen presentation of the data in conjunction with the proportional scaling of $P$ and $B_{\rm bias}$ by the factor $\kappa$.}
	\label{fig:TrapDepth} 
\end{figure}

We now investigate the dependence of the optical wire trap on various parameters that can be readily varied experimentally. The distance between the trap minimum and the fiber surface can be influenced by altering the ratio of the strengths of the fictitious field and the bias field. In Fig.~\ref{fig:TrapPos}, the trapping potential $U(x=0,y)$ is shown for a fixed optical power $P=1.2$~mW and different strengths of the bias field in the range $B_{\rm bias}=10, \ldots ,34$~G for increments of $3$~G. For each calculated potential, we add an energy offset such that $U(x=0,y \rightarrow \infty)=0$. For increasing $B_{\rm bias}$, the separation between the trap minimum and the surface of the fiber reduces. For the trap configurations with $B_{\rm bias}=\{25,~28,~31,~34\}$~G, the relative contribution of $U_{\rm vdW}$ to the total trapping potential is so large that the potential opens towards the surface of the nanofiber. Changing the trap depth while keeping the position of the trap minimum approximately constant can be achieved by scaling $P$ and $B_{\rm bias}$ proportionally. In Fig.~\ref{fig:TrapDepth}, the trapping potential $U(x=0,y)$ is shown for $P=\kappa\cdot1.2~{\rm mW}$ and $B_{\rm bias}=\kappa\cdot22~{\rm G}$ with $\kappa=\{0.5,~1,~1.5,~2,~2.5,~3\}$.

Table~\ref{tab:parameters} lists the trap depth $U_0$, the separation $y_0-a$ between the trap minimum and the surface, and the radial and azimuthal trap frequencies  ($\omega_r$,~$\omega_\varphi$) for five example trapping configurations. The traps (a) to (d) have a depth that is suitable for storing laser-cooled atoms and exhibit trapping frequencies between a few ten kHz and a few hundred kHz. Configuration (e) exhibits a smaller trap depth, aimed for the storage of ultracold atoms. This configuration requires a lower trapping laser power of only $P=150~\mu$W and yields trap frequencies of a few ten kHz. In all five cases, the separation between the trap minimum and the surface of the fiber is between about 100~nm and 200~nm.

A small distance between the trap minimum and the surface of the nanofiber allows one to efficiently interface the atoms with additional guided light fields. A simple and common measure for the strength of this coupling is given by the resonant optical depth per atom $\sigma/A_{\rm eff}$~\cite{Domokos02,Vetsch10,Goban12}. Generally, the on-resonance cross section of the atom is given by $\sigma=\hbar \omega \Gamma /(2 I_{\rm sat})$ with the decay rate of the excited state $\Gamma$ and the saturation intensity $I_{\rm sat}$. Here, we approximate $\sigma$ as $\sigma_0=3 \lambda^2/2\pi$, which is the on-resonance cross section of the atom driven on a cycling transition in free space. We note that this description neglects a possible increase of the spontaneous emission rate of the trapped atom due its close proximity to the fiber surface~\cite{LeKien05c,LeKien06c,LeKien08c}. The effective mode area $A_{\rm eff}$ of a guided probe field at the trap minimum $(x=0,y_0)$ is given by $A_{\rm eff}=P_{\rm probe}/I_{\rm probe}$, with $P_{\rm probe}$ being the power of the guided probe field and $I_{\rm probe}=c \epsilon_0 \langle |\mathbf{E}(t)|^2 \rangle_T$ the intensity of the probe field at the trap minimum. In the definition of the intensity, $c$ is the speed of light in vacuum, $\epsilon_0$ is the electric permittivity, and $\langle \ldots \rangle$ denotes the time average of the square of the absolute value of the electric field over one oscillation period. In Tab.~\ref{tab:parameters}, $\sigma_0/A_{\rm eff}$ is evaluated for a quasi-linearly-polarized probe light field with its plane of polarization orientated such that the intensity at the position of the atoms is maximized~\cite{LeKien04b}, i.e., with its main plane of polarization containing the fiber axis and the trap minima. The field is chosen to be resonant with the cesium D2 cycling transition $F=4  \to F'=5$ at a wavelength of $852$~nm. All configurations show large values of $\sigma_0/A_{\rm eff}$, with the weakest coupling being 0.17 (e) and the strongest 0.43 (c). This significantly exceeds the optical depth per atom of 0.08 that has been obtained experimentally with the two-color nanofiber-based trapping scheme~\cite{Goban12} and is a result of the smaller trap--surface distance that can be achieved in the case of the optical wire trap.
\begin{table*}
		\begin{tabular}{|c||c|c|c|c|c|c|c|c|c|}
			\hline
			 & $P$ (mW) & $B_{\rm bias}$ (G) & $U_0/k_{\rm B}$ ($\mu$K) & $y_0-a$ (nm) & $\omega_r/(2\pi)$ (kHz) & $\omega_\varphi/(2\pi)$ (kHz) & $\sigma_0/A_{\rm eff}$ & $\Gamma_{\rm sf}$ ($s^{-1}$) & $\Gamma_{\rm exc}$ ($s^{-1}$)
		 \\
			\hline \hline
			(a) & 1.2 & 16 & 210 & 189 & 247 & 67 & 0.20 & $8\times10^{-7}$ & 16.0 \\ \hline
			(b) & 1.2 & 22 & 310 & 148 & 307 & 92 & 0.30 & $7\times10^{-8}$ & 23.4 \\ \hline
			(c) & 1.2 & 28 & 150 & 115 & 357 & 119 & 0.43 & $6\times10^{-9}$ & 32.1 \\ \hline
			(d) & 2.4 & 44 & 608 & 150 & 433 & 128 & 0.30 & $2\times10^{-13}$ & 46.0 \\ \hline
			(e)$^\ast$ & 0.15 & 2.6 & 7.1 & 205 & 45 & 19 & 0.17 & $1.5\times10^{-9}$ & 1.7 \\ \hline
		\end{tabular}
		\caption{Example configurations of the optical wire trap and respective trap parameters: $P$ optical power of the guided trapping field, $B_{\rm bias}$ magnetic bias field, $U_0/k_{\rm B}$ trap depth, $y_0-a$ separation between trap minimum and nanofiber surface, $\omega_r$ radial trap frequency, $\omega_\varphi$ azimuthal trap frequency, $\sigma_0/A_{\rm eff}$ optical depth per atom, $\Gamma_{\rm sf}$ spin flip rate, $\Gamma_{\rm exc}$ trapping-light-induced excitation rate. The $^\ast$ indicates the presence of an additional magnetic offset field along $-z$ of 1~G.}
		\label{tab:parameters}
\end{table*}

Cold atoms trapped in an inhomogeneous magnetic field can undergo spin flips~\cite{Sukumar97} when their magnetic moment does not adiabatically follow the orientation of the local magnetic field. An atom can, for example, undergo a transition from a trapped (low-field-seeking) state to an untrapped (magnetic-field-insensitive or high-field-seeking) state and be lost from the trap. This effect becomes significant if atoms move through regions of small or even vanishing magnetic field, e.g., in a linear or spherical magnetic quadrupole field. Spin flips can be efficiently suppressed by an offset magnetic field or by using time-averaged potentials. In order to quantify the effect of spin flips in our system, we calculate the spin flip rate $\Gamma_{\rm sf}$ for the trapping configurations (a) to (e) using the formula~\cite{Sukumar97,Brink06}
\begin{eqnarray} \label{eq:spinflips}
\Gamma_{\rm sf}=\frac{\pi \omega_t}{2} \exp \left(- \frac{\pi E_0}{2 \hbar \omega_t} \right),
\end{eqnarray}
where $\omega_t$ is the trap frequency, $\hbar$ is the reduced Planck constant, and $E_0=\mu_B |g_{nJF}| |\mathbf{B}|$ is the energy gap between the potential branches for the different states. For all example configurations, we obtain negligible spin flip rates as summarized in Tab.~\ref{tab:parameters}. In order to perform a conservative estimation, the larger of the two trap frequencies, $\omega_r$, has been used for the calculations. Advantageously, the axial component of the fictitious magnetic field shown in Fig.~\ref{fig:principle} acts as an integrated offset field for the optical wire trap, thereby intrinsically suppressing spin flips. This intrinsic offset field can be changed by varying the fiber radius $a$, thereby changing the local polarization of the guided trapping laser field and, thus, the ratio between the azimuthal and axial components of $\mathbf{B}^{\mathrm{fict}}$. Additionally, the offset field can be modified using an axial component of the external magnetic field as it is typically done for conventional magnetic wire traps. This method was applied for example (e) in Tab.~\ref{tab:parameters}.

An atom trapped in state $\ket{F=4,m_{\rm F}=4}$ may also undergo a change of its hyperfine- and/or Zeeman state due to spontaneous Raman scattering of trapping light. In Tab.~\ref{tab:parameters}, we calculate the light-induced excitation rate $\Gamma_{\rm exc}$ according to \cite{InPreparation}
\begin{eqnarray}
\Gamma_{\rm exc} = I_{\rm trap} \left[ \eta_s + (-1)^{F-I+1/2} m_F C \eta_v \right],
\end{eqnarray}
with $I_{\rm trap}(x=0,y_0)$ being the intensity of the trapping field at the trap minimum, $\eta_s=0.2446$~kHz cm$^2$/MW and $\eta_v=2.860\cdot10^{-2}$~kHz cm$^2$/MW the scalar and vector scattering coefficients, $F$ and $I$ the conventional quantum numbers of the total spin and the spin of the atomic core, respectively, and $C$ the trapping field polarization ellipticity~\cite{LeKien13a}. The rate $\Gamma_{\rm exc}$ is a worst-case estimate as spontaneous scattering does not necessarily change the atomic state. However, we expect inelastic scattering to be enhanced because the wavelength of the optical wire trap lies in between the two D-lines~\cite{Cline94}.  The values obtained for the configurations (a) to (e) are on the order of $1~{\rm s}^{-1}$ to $10~{\rm s}^{-1}$ which, despite the smaller detuning of the trapping field, is comparable to the rates for the two-color trap realized in \cite{Vetsch10}. This is due to the low intensity of the trapping light at the position of the atoms as compared to the two-color scheme. Note that optical-wire-trapping of other alkali atoms than cesium should be feasible, too. However, for the same trap depths, the scattering rate will be larger by a factor of about $3$, $12$, and $67$ for rubidium, potassium, and sodium, respectively~\cite{Cho97}.

The potentials presented above are translationally invariant along the fiber and thus form a guide for atoms parallel to the fiber axis. Additional axial confinement can be straightforwardly provided by means of an externally applied inhomogeneous magnetic field. Moreover, counter-propagating fiber-guided fields can yield a periodic axial modulation of the fictitious magnetic field~\cite{ArXiv_LeKien13c} and, thus, might allow one to form a periodic array of trapping sites.

In summary, we proposed a novel nanofiber-based trapping scheme and analyzed important trap parameters in detail. Our optical wire trap features depths which are sufficient to store laser-cooled atoms while using optical powers of the trapping field which are one order of magnitude smaller than these for a typical two-color nanofiber-based trap. Thanks to the close analogy of our scheme to conventional wire traps for paramagnetic atoms, established techniques from this field such as the adiabatic transformation of trapping potentials for trap loading, evaporative cooling of the atoms, or the interrogation of the atomic ensemble using dispersive light fields should be applicable for the optical wire trap~\cite{Fortagh98,Ketterle99,Miller02,Dawkins11}. Our scheme can access trap parameters comparable to the steepest wire traps formed close to miniaturized wires~\cite{Reichel02,Fortagh07} such as current-carrying carbon nanotubes~\cite{Peano05,Salem10,Fermani07}, thereby opening up opportunities for studying surface physics~\cite{Klimchitskaya09} and one-dimensional matter wave dynamics~\cite{Petrov00,Haller10}. Additional magnetic trapping configurations might arise due to the fact that fictitious magnetic fields can also have a local maximum in free space which is not possible for real magnetic fields~\cite{Wing84}. In contrast to conventional wire traps, surface-induced spin flips due to Johnson noise~\cite{Henkel99} will be negligible in our system~\cite{Reitz13}, even for small separations of the atoms to the dielectric surface of the nanofiber. For typical configurations of the optical wire trap, the local minima of the potential are about 100 to 200~nm away from the fiber surface which is a key advantage when it comes to optically interfacing trapped atoms with further nanofiber-guided light fields. In our system, optical depths per atom of up to about 0.4 are accessible, opening a realm of applications in nanofiber-based nonlinear and quantum optics. 

We thank R.~Grimm and J.~Schmiedmayer for helpful comments and discussions. Financial support by the Wolfgang Pauli Institute and the Austrian Science Fund (FWF; Lise Meitner project No. M 1501-N27 and SFB NextLite project No. F 4908-N23) is gratefully acknowledged.


\bibliography{Vector}

\end{document}